\documentclass[sigconf,natbib=true]{acmart}
\usepackage{tikz}
\usepackage{bbold}
\usepackage{ bm        }
\usepackage{ amssymb   }
\usepackage{ mathdots  }
\usepackage{ amsmath   }
\usepackage{ color     }
\usepackage{ graphicx  }
\usepackage{ amsfonts  }
\usepackage{ amssymb   }
\usepackage{ epstopdf  }
\usepackage{ upgreek   }
\usepackage{ bm        }
\usepackage{ graphicx  }
\usepackage{ subfigure }
\usepackage{ mathtools }
\usepackage{ algorithm }
\usepackage{algorithmic}
\usepackage{ booktabs  }
\usepackage{ multirow  }
\usepackage{ caption   }
\usepackage{ array     }
\usepackage{ float     }
\usepackage{   soul    }
\usepackage{adjustbox}
\usepackage{newfloat}
\usepackage{listings}

\AtBeginDocument{%
	\providecommand\BibTeX{{%
			\normalfont B\kern-0.5em{\scshape i\kern-0.25em b}\kern-0.8em\TeX}}}

\acmConference[DLP-KDD 2022]{4th Workshop on Deep Learning Practice and Theory for High-Dimensional Sparse and Imbalanced Data with KDD 2022}{August 14, 2022}{Virtual Conference, Online}
\acmPrice{15.00}
\acmISBN{978-1-4503-XXXX-X/22/08}

\begin{document}
	
	\title{Efficient Long Sequential User Data Modeling for Click-Through Rate Prediction}
	
	\author{Qiwei Chen}
	\email{chenqiwei.cqw@alibaba-inc.com}
	\affiliation{%
		\institution{Alibaba Group.}
		\state{}
		\country{}
	}
	
	\author{Yue Xu}
	\email{yuexu.xy@foxmail.com}
	\affiliation{%
		\institution{Alibaba Group.}
		\state{}
		\country{}
	}
	
	\author{Changhua Pei}
	\email{changhuapei@gmail.com}
	\affiliation{%
		\institution{Alibaba Group.}
		\state{}
		\country{}
	}
	
	\author{Shanshan Lv}
	\email{lss271346@alibaba-inc.com}
	\affiliation{%
		\institution{Alibaba Group.}
		\state{}
		\country{}}
	
	\author{Tao Zhuang}
	\email{zhuangtao.zt@alibaba-inc.com}
	\affiliation{%
		\institution{Alibaba Group.}
		\state{}
		\country{}}
	
	\author{Junfeng Ge}
	\email{beili.gjf@alibaba-inc.com}
	\affiliation{%
		\institution{Alibaba Group.}
		\state{}
		\country{}}
	

	\begin{abstract}
		Recent studies on Click-Through Rate (CTR) prediction has reached new levels by modeling longer user behavior sequences. 
		Among others, the two-stage methods stand out as the state-of-the-art~(SOTA) solution for industrial applications. The two-stage methods first train a retrieval model to truncate the long behavior sequence beforehand and then use the truncated sequences to train a CTR model. However, the retrieval model and the CTR model are trained separately. So the retrieved subsequences in the CTR model is  inaccurate, which degrades the final performance.
		In this paper, we propose an end-to-end paradigm to model long behavior sequences, which is able to achieve superior performance along with remarkable cost-efficiency compared to existing models.
		Our contribution is three-fold:
		First, we propose a hashing-based efficient target attention~(TA) network named ETA-Net to enable end-to-end user behavior retrieval based on low-cost bit-wise operations. The proposed ETA-Net can reduce the complexity of standard TA by orders of magnitude for sequential data modeling.
		Second, we propose a general system architecture as one viable solution to deploy ETA-Net on industrial systems. Particularly, ETA-Net has been deployed on the recommender system of Taobao, and brought $ 1.8\% $ lift on CTR and $ 3.1\% $ lift on Gross Merchandise Value~(GMV) compared to the SOTA two-stage methods.
		Third, we conduct extensive experiments on both offline datasets and online A/B test. The results verify that the proposed model outperforms existing CTR models considerably, in terms of both CTR prediction performance and online cost-efficiency.
		ETA-Net now serves the main traffic of Taobao, delivering services to hundreds of millions of users towards billions of items every day.
	\end{abstract}
	\maketitle

	\section{Introduction}
	Click-Through Rate (CTR) prediction plays an important role in industrial recommender systems (RS). 
	In the past decade, user behavior sequences have been widely recognized to be of great potential in boosting the performance of CTR prediction~\cite{zhou2018deep,chen2019behavior,ren2019lifelong,pi2019practice,zhou2019deep,qi2020search,qin2020user}. This is due to the fact that they contain valuable sequential patterns that are beneficial for capturing a user's evolving interests.
	Such benefit has therefore attracted extensive research interest on user sequence modeling. 
	Increasing the length of user behavior sequences in a CTR model is  beneficial since it provides more long-term behavior patterns~\cite{ren2019lifelong}. 
	And recent studies start to explore the feasibility of modeling long sequences~(with length up to $ 1000 $ or more) and achieved positive results~\cite{zhou2019deep,qi2020search,qin2020user}.
	Nonetheless, processing long sequences in a CTR model takes a lot more computation. In an industrial recommender system which has strict online latency and memory limit, it remains a great challenge to utilize long sequences in CTR prediction more efficiently. 
	
	Generally, the most challenging part of long sequential data modeling is how to effectively extract the most valuable information from long sequences and feed it into the CTR model for later inference.
	The SIM~\cite{qi2020search} and UBR4CTR~\cite{qin2020user} models, which has achieved  state-of-the-art~(SOTA) performances in industrial applications, adopted a two-stage paradigm to model long sequences.
	The first stage  retrieves the top-$k$ items relevant to the user from his long behavior sequences using a pre-trained retrieval model. And the results are cached in an  database. The second stage predict the user's CTR using the top-$k$ items relevant to the user.  
	However, in the two-stage paradigm, the retrieval model and the CTR model are trained separately. Therefore, the user and item representations used in different stages are not updated synchronously. This will degrade the prediction performance of the CTR model. 
	Moreover, in order to meet the online latency and memory requirements, the SIM model takes the category of an item, instead of the item ID, as the retrieval key. Approximating an item with its category attribute would lead to further performance loss.
	And this approximation is not applicability to other recommendation tasks where such discriminative attributes like item category does not exist. For example, in short video recommendations, the videos are usually classified by a series of tags instead of an explicit category.
	
	In this paper, we propose the Efficient Target Attention Network (ETA-Net) for long user behavior sequence modeling. ETA-Net aims to efficiently retrieve valuable information from long sequences in a more principled manner under acceptable complexity for large-scale industrial applications. ETA-Net performs user behavior retrieval based on low-cost bit-wise operations. 
	We design a general system architecture in order to guide the deployment of the proposed ETA-Net to large-scale industrial systems. ETA-Net has been deployed on the RS of Taobao, and brought $ 1.8\% $ lift on  online CTR and $ 3.1\% $ lift on Gross Merchandise Value~(GMV) which amounts to  millions of clicks and tens millions of GMV increase every day.
	It now serves as the primary CTR model for Taobao recommendation system.
	The main contributions are summarized as follows.
	
	\begin{itemize}
		\item \textbf{Algorithm:} We propose the ETA-Net as a new paradigm for efficient long sequence modeling. Particularly, we propose a hashing-based efficient target attention~(TA) algorithm which replaces the dot-product operations in the standard TA with low-cost bit-wise operations. This reduces the computational complexity by orders of magnitude. To the best of our knowledge, it is the first user interest model which is able to model long sequence in a principled end-to-end manner.
		
		\item \textbf{Architecture:} We present a general system architecture as one viable solution to deploy ETA-Net in industrial recommender systems. The proposed architecture has been implemented in the online RS of Taobao and is able to serve the main traffic with around $ 120,000 $ queries-per-second~(QPS) at peak. The online deployment of ETA-Net brings 1.8\% lift on CTR and 3.1\% lift on GMV compared to the SOTA two-stage methods.
		
		\item \textbf{Verification:} We conduct extensive experiments on both offline datasets and online A/B test. The results verify that our proposed ETA-Net significantly outperforms the SOTA models. We also provide ablation studies to analyse the specific benefits of  long sequence modeling.
	\end{itemize}
	
	
	\section{Related Work}
	Deep learning based CTR models have achieved remarkable performances by making good use of the user and item features~\cite{cheng2016wide,wang2017deep,guo2017deepfm,qu2016product,lian2018xdeepfm}. 
	In recent years, researchers start to capture the dynamic shift of users' interest by modeling user behavior sequences, which gives rise to many new benchmarks in CTR prediction~~\cite{zhou2018deep,chen2019behavior,ren2019lifelong,pi2019practice,zhou2019deep,qi2020search,qin2020user}. 
	For example, DIN~\cite{zhou2018deep} proposed an attention based method named TA to capture the diverse interests of a given user against different items. 
	DIEN~\cite{zhou2019deep} proposed an interest extraction layer based on Gated Recurrent Unit~(GRU) to model users' drifting temporal interest from their historical behaviors. 
	MIND~\cite{li2019multi} proposed to model users' diverse interest with multi-vectors instead of a single vector based on dynamic routing and the Capsule network~\cite{sabour2017dynamic}.
	Though effective, these models all focus on the modeling of users' short sequences, typically less than $ 100 $, due to the fast growing complexity along with the user sequence length. This limitation restricts their performance to some extend. 
	
	In this context, recent studies start to investigate how to make use of longer user behavior sequences for performance enhancement. 
	Among others, the following two paradigms have been verified to be effective and applicable to online industrial systems.  
	The first is the memory-based methods which proposed to model long sequences using a memory network. 
	For example, HPMN~\cite{ren2019lifelong} and MIMN~\cite{pi2019practice} proposed to decouple user interest modeling from CTR prediction by maintaining the diverse interest of one user into a fixed size memory matrix.  
	However, it is still challenging for these approaches to model extreme long sequential data due to the memory limit and the inherent noise~\cite{qi2020search}.
	
	The other alternative is the two-stage methods which surpasses the memory-based methods to be the SOTA solution, such as SIM~\cite{qi2020search} and UBR4CTR~\cite{qin2020user}.
	These two-stage methods propose to train a retrieval model and a CTR model separately.
	Specifically, at the first stage, the retrieval model retrieves the top-$k$ relevant items from long user behavior sequences and store the subsequence in an offline database according to the retrieval key. The retrieval key are determined by user and item representations simultaneously.
	Then, at the second stage, the CTR model retrieves the top-$k$ relevant items directly from the offline database by referring to the retrieval keys. 
	However, the user and item representations are evolving during the training of the CTR model at the second stage, such that the prestored top-$k$ items at the first stage can be inaccurate. This mismatch may downgrade the final performance.
	Moreover, when deploying SIM to the online RSs, the authors proposed to directly employ user ID and item category as the retrieval key in order to meet the delay and memory limits.
	This approximation may leads to extra performance loss and limits its applicability to other recommendation tasks where such discriminative attributes like item category may not exist, e.g., video recommendation. 
		The above limits motivate us to explore a more principled and general paradigm for long sequence modeling.
	
	Noticeably, recent advances on efficient Transformers~\cite{tay2020efficient} has achieved remarkable progress in natural language processing~(NLP) for long sequence modeling.
	Particularly, Reformer~\cite{kitaev2020reformer} proposed to replace the dot-product attention in Transformer with a locality-sensitive hashing~(LSH) attention so as to largely reduce the complexity of relevant item retrievals.
	However, Reformer can not be directly applied to the recommendation tasks due to the violation of their assumptions on data structures. More details are provided in Sec.~\ref{sec:methodology}.  
	
	\section{Problem Definition}
	\label{sec:problem_definition}
	CTR prediction is usually modeled as a binary classification problem.
	Specifically, given an impression~$m$ where an item~$ i $ is impressed to an user~$ u $, the goal is to predict the probability of target user~$ u $ clicking the target item~$ i $ under a given feature set~$\mathcal{X}_m$, which can be written as
	\begin{equation}
		p^{m} = P(y^m=1|~\mathcal{X}^m; \theta), \quad \forall \textit{m} \in \mathcal{D},
	\end{equation}
	where $y^{m}$ refers to the binary label which is marked as one if click happens otherwise zero, $\mathcal{X}^m$ denotes the input features of the CTR model at impression~$ m $, $ \theta $ denotes the trainable parameters and $ \mathcal{D} $ is the entire dataset.
	The input feature $\mathcal{X}^m$ in recent studies usually contains five categories: the user profile features $\mathcal{X}^m_u$, the context features $\mathcal{X}^m_c$, the item attributes $\mathcal{X}^m_i$, the short user behavior sequences $ \mathcal{X}^m_{st} $ and the long user behavior sequences $ \mathcal{X}^m_{lt} $. 
	The common routine to model $\mathcal{X}^m_u$, $\mathcal{X}^m_c$ and $\mathcal{X}^m_i$ is to feed them into multiple multi-layer perceptron~(MLP) layers~\cite{cheng2016wide,guo2017deepfm,zhou2018deep,pi2019practice,zhou2019deep,qi2020search,qin2020user}.
	While the modeling of $ \mathcal{X}^m_{st} $ is usually based on the TA structure proposed in DIN~\cite{zhou2018deep}. 
	
	However, the computational complexity of standard TA scales as $\mathcal{O}(L \cdot d)$ where $L$ is the size of $ \mathcal{X}^m_{lt} $ and $d$ is the size of the embedding vector. In other words, the complexity grows linearly with the length of user behavior sequence.
	Moreover, for each user request in online RS, the CTR model needs to perform CTR prediction over multiple candidate items, i.e., $ \forall i \in \mathcal{I}^c_u $ where $ \mathcal{I}_c $ denotes the set of candidate items for the user request.
	In this case, the long sequence modeling based on standard TA  incurs a complexity of $\mathcal{O}(N_c \cdot L \cdot d)$ per user request, where $ N_c $ is the size of $ \mathcal{I}_c $. This indicates that the system latency increases linearly~(or worse) with both $L$ and $d$, which largely hinders standard TA to be applied to model long sequences in industrial applications.
	
	Therefore, in this paper, we aim to 
	1)~propose a low-cost TA algorithm that is able to model the long behavior sequences $ \mathcal{X}_{lt} $ with much lower complexity than $\mathcal{O}(L \cdot d)$;
	2)~present a system architecture to envision how to implement the proposed efficient algorithm in industrial systems with billion-scale user and items;
	3)~verify the effectiveness of the proposed algorithm and architecture through online and offline experiments.
	
	\section{Methodology}
	\label{sec:methodology}
	In this section, we start with an overview of our proposed ETA-Net and then introduce the design of ETA in detail. We also present a general system architecture to envision how to implement ETA-Net in real-world RSs along with our hands-on experiences on the deployment of ETA-Net in Taobao.
	
	\subsection{Model Overview}
	\begin{figure*}[tb]
		\centering
		\includegraphics[trim = 30 10 20 10, clip, width=16cm]{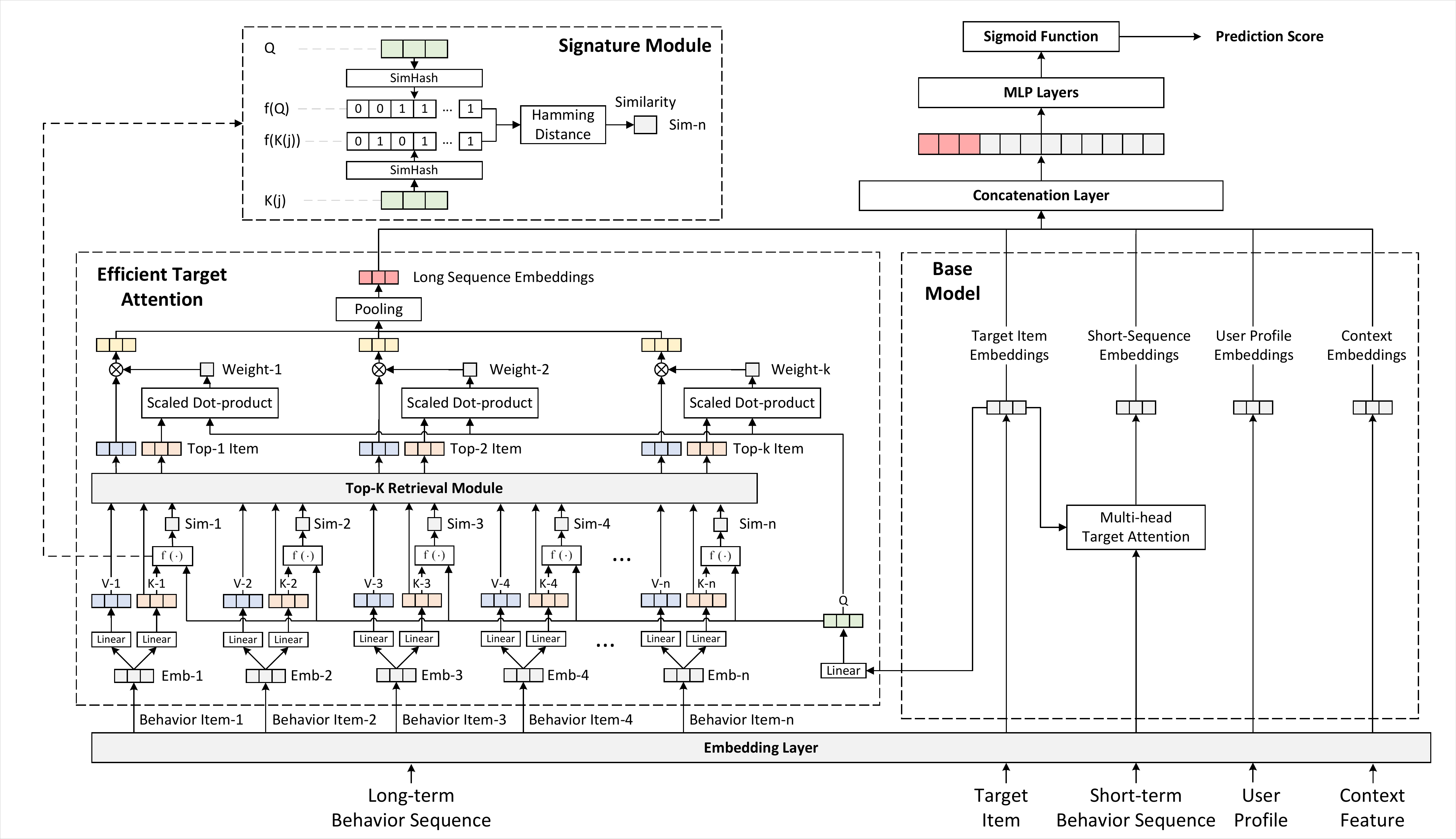}
		\caption{The illustration of our ETA-Net (Efficient Target Attention Network). ETA-NET is composed of two main parts: (i) the ETA part focuses on modeling user long term interest from over few thousands user behaviors, (ii) the other part follows the traditional Embedding\&MLP paradigm which takes the output of ETA and other features as inputs. The contribution of ETA-NET lies in the ETA part, ETA generates a binary fingerprint for each behavior item $ i' \in \mathcal{X}^m_{lt} $ and target item $\mathcal{X}^m_i$ based on the SimHash method, and selects the top-$ K $ relevant behavior items to the target item by comparing the hamming distance (a low-cost bit-wise operations) of their fingerprints, then utilizes multi-head attention to capture the diverse user interest from most similar K behavior.}
		\label{fig:overview}
	\end{figure*}
	
	As shown in Figure~\ref{fig:overview}, ETA-Net takes user-side and item-side features as input to predict the click probability of a target user $ u $ towards a target item $ i $. 
	Generally, ETA-Net consists of two parts, i.e., the ETA part to model the long sequences and the base model to model the other features.  
	The base model processes user features~$\mathcal{X}^m_u$, item features~$\mathcal{X}^m_i$ and context features~$\mathcal{X}^m_c$ based on multiple MLP layers and process short sequences~$\mathcal{X}^m_{st}$ based on standard TA.

	The ETA part processes long behavior sequences $ \mathcal{X}^m_{lt} $ based on our proposed ETA algorithm.
	Specifically, ETA first generates a binary fingerprint for each behavior item $ i' \in \mathcal{X}^m_{lt} $ and the target item $\mathcal{X}^m_i$ based on the SimHash method~\cite{charikar2002similarity}; then, ETA selects the top-$ K $ relevant behavior items to the target item by comparing the hamming distance of their fingerprints, which can be readily calculated via low-cost bit-wise operations; next, ETA uses the top-$ K $ relevant items to model user's long-term interest.
	In general, ETA-Net is able to reduce the computational complexity of standard TA from $\mathcal{O}(L \cdot d)$ to $\mathcal{O}(L) + \mathcal{O}(K \cdot d)$ with $K \ll L $ when modeling sequential user data.  
	Also note that the ETA part can be jointly trained with the base model in an end-to-end manner. This distinguishes it from the two-stage methods such as SIM, where the retrieval model and the CTR model are trained separately.
	
	
	\subsection{Efficient Target Attention Network}
	\label{sec:eu}
	In this section, we introduce the proposed ETA-Net in detail. Specifically, we start by introducing the standard dot-product based TA as preliminaries. Then, we introduce our proposed bit-operation based ETA with comparisons to standard TA and Reformer.
	
	\subsubsection{Dot-product Target Attention.}
	\label{sec:mhta}
	Standard TA is based on the structure of multi-head attention~(MHA)\cite{vaswani2017attention} and has been widely applied to sequential user data modeling in CTR models~\cite{zhou2020can,xiao2017attentional,qin2020user,qi2020search,xu2020deep}. 
	Different with the MHA in NLP models, the MHA in CTR models treats the target item as query and treats the user behavior sequence as key and value. As such, the query in MHA only consists of a single item instead of a sequence of items. We refer to this MHA structure as multi-head target attention~(MHTA) in sequel.
	
	The formulation of one single-head attention in MHTA can be written as
	\begin{equation}\label{eq:self_att}
		\text{Att}(\bm{Q,K,V}) = \text{Softmax}\left(\alpha \bm{QK}^T \right)\bm{V},
	\end{equation}
	where $ \bm{Q} \!=\! \bm{E}^t(\bm{W}_i^Q)^T $, $ \bm{K} \!=\! \bm{E}^s(\bm{W}_i^K)^T $ and $ \bm{V} \!=\! \bm{E}^s(\bm{W}_i^V)^T $ are linear transformations applied to the input embeddings.
	Here $\bm{E}^t \in \mathbb{R}^{1 \times d_q}$ and $ \bm{E}^s \in \mathbb{R}^{L \times d_k}$ denote the embedding matrices of the target item and the behavior items, respectively; $ L $ denotes the length of behavior sequences; $d_q$ and $d_v$ denote the embedding size of the target item and the behavior items, respectively. $\bm{W}_i^Q \in \mathbb{R}^{d \times d_q}$, $\bm{W}_i^K \in \mathbb{R}^{d \times d_k}$ and $\bm{W}_i^V \in \mathbb{R}^{d \times d_v}$ denote the linear projection matrices. The scaling factor $ \alpha $ is usually set to be $ 1/\sqrt{d_k} $.
	In this context, the full MHTA can be written as
	\begin{subequations}\label{eq:dot_ta}
		\begin{align}
			\text{Head}_i &= \text{Att}(\bm{Q}, \bm{K}, \bm{V}), \\
			\text{TA}(\bm{E}^t,\bm{E}^s) &= \text{Concat}(\text{Head}_1,\cdots,\text{Head}_h)\bm{W}^O,  
		\end{align}
	\end{subequations}
	where $ \text{Concat}(\cdot) $ denotes the concatenation operation and $\bm{W^O} \in \mathbb{R}^{(n_{H}\cdot d_v) \times d}$ denotes the linear projection matrix with $n_H$ denoting the number of used heads.
	

	As shown in (\ref{eq:dot_ta}), MHTA needs to compute \mbox{$ \bm{Q}\bm{K}^T $} at each head, which incurs a complexity of $\mathcal{O}(L \cdot d_k)$ when modeling a sequence of length $ L $. 
	As stated in Sec.~\ref{sec:problem_definition}, in online RS, the model needs to score a list of candidate items in order to return the top items to users per request, which will further increase the complexity from $\mathcal{O}(L \cdot d_k)$ to $\mathcal{O}(N_c \cdot L \cdot d)$.
	Considering that the embedding size $d$ and the number of candidate items $N_c$ usually range from a few hundreds to a few thousands, the RS has to cut the length of the behavior sequence, i.e., $L$, so as to meet the latency requirement~\cite{pi2019practice}. For example, DIN and DIEN restrict $ L $ to be less than~$ 100 $~~\cite{zhou2018deep,zhou2019deep}.
	However, such strict restriction on the length of user behavior sequences will inevitably downgrade the prediction performance. This motivates researchers to investigate a more efficient solution to deal with long sequence modeling in recommendation tasks.
	
	\subsubsection{Efficient Target Attention.}
	\label{sec:ETA}
	The computational bottleneck of standard TA lies on the computation of~$ \bm{Q}\bm{K}^T $ at each head. 
	However, the purpose of evaluating $ \bm{Q}\bm{K}^T $ is to measure the similarity between the target item and the behavior items. As such, what really matters is the output of $ \text{Softmax} \big( \bm{Q}\bm{K}^T \big) $ instead of $ \bm{Q}\bm{K}^T $.
	Since the output of $ \text{Softmax} \big( \bm{Q}\bm{K}^T \big) $ is dominated by the largest elements of $ \bm{Q}\bm{K}^T $, i.e., the most similar elements between $ \bm{Q}$ and $\bm{K} $, we can turn the problem of how to evaluate $ \bm{Q}\bm{K}^T $ efficiently into \textit{how to retrieve the top-$k$ behavior items that are most similar with the (projected) target item in an efficient way}.
	
	The recently proposed LSH attention in Reformer~\cite{kitaev2020reformer} presents a promising solution to efficiently retrieve the most similar items from long sequences and has been applied to many NLP tasks. 
	However, LSH attention poses two assumptions on the input data:
	1)~the query and the key are equivalent, i.e., $ \bm{Q} = \bm{K} $;
	2)~the query and the key have been sorted according to their hash buckets since similar items fall in the same bucket with high probability.
	Nevertheless, the above assumptions are not applicable to CTR prediction due to the fact that 1)~the target item and the behavior items are different and 2)~ sorting hundreds or thousands of behavior items per user request are costly.
	Therefore, LSH attention are not directly applicable to CTR models. 
	Inspired by Reformer, we herein propose a SimHash-based attention structure named ETA which not only allows the query and the key to be different but also enables fast retrieval of similar items based on low-cost bit-wise operations, which is able to deal-with large-scale online recommendation tasks.
	
	Specifically, the output of one particular head $ h $ in standard TA can be reformulated as
	\begin{equation}\label{key}
		H_h = \sum_{j\in \mathcal{B}} \exp\Big( \bm{Q} \cdot \bm{K}_j - z(\mathcal{B}) \Big) \bm{V},
	\end{equation}
	where $ \mathcal{B} $ denotes the set of behavior items attend to and $ z $ denotes the normalizing term in the softmax function with the scaling factor $ \sqrt{d_k} $ omitted. As aforementioned, instead of attending target item to all behavior items, we only need to attend target item to its most similar behavior items, which generates the following restriction:
	\begin{equation}\label{eq:restriction}
		\mathcal{B} = \Big\{ j: h(\bm{Q}) = h(\bm{K}_j) \Big\}.
	\end{equation}
	Here $ h(\cdot) $ refers to a hashing function that uses an N-nary fingerprint to encode an vector. 
	The hashing function must satisfy the Locality-sensitive properties~\cite{kitaev2020reformer}, i.e., the output hash codes are similar if the input vectors are similar to each other. 
	Different from LSH attention, we adopt \textit{binary fingerprints} instead of multi-nary fingerprints to represent the embedding vectors.
	In this way, ETA is able to measure the similarity between the target item and the behaviors by comparing their binary fingerprints through low-cost bit-wise operations. 
	
	Specifically, the formulation to obtain one $ m $-bit fingerprints based on SimHash can be written as
	\begin{equation}\label{key}
		h(\bm{e}) = \mathbb{1}_{\mathcal{R}^-} ( \bm{e} \cdot \bm{H} ),
	\end{equation}
	where $ \bm{e} \in \mathbb{R}^{1\times d} $ denotes the embedding vector to encode, $ \bm{H} \in \mathbb{R}^{d\times m} $ is a random hash matrix where each column represents one distinct hash function $ h_k$ with $k \in \{1,2,\cdots, m\} $, and $ \mathbb{1}_{\mathcal{R}^-} (x)$ denotes a binary indicator function which equals to zero if $ x < 0 $ otherwise one.
	
	On the other hand, it is possible that similar items fall into different buckets under one round of hashing encoding. 
	In order to reduce this probability, ETA performs multiple rounds of hashing so as to generate multiple sets of binary fingerprints of $ \bm{Q} $ and $ \bm{K} $ and measure their similarity based on the Hamming distance of all sets of fingerprints, similar as Reformer. 
	Specifically, such multi-round hash encoding turns the restriction in (\ref{eq:restriction}) into
	\begin{equation}\label{eq:restriction2}
		\mathcal{B} \!=\! \cup^{n_r}_{r=1} \mathcal{B}^{(r)}, ~~\mathcal{B}^{(r)} \!=\! \Big\{ j: h^{(r)}(\bm{Q}) = h^{(r)}(\bm{K}) \Big\},
	\end{equation}
	where $n_r$ is the number of repeated rounds and $ h^r $ denotes the hashing function at round $ r $.
	ETA measures the similarity of $ \bm{Q} $ and $ \bm{K} $ based on the hamming distance of $ h(\bm{Q})$ and $ h(\bm{K})$. 
	Note that both $ h(\bm{Q})$ and $ h(\bm{K})$ are binary vectors, such that their hamming distance can be readily measured through bit-wise $ xor $ operation, which incurs constant complexity~$ \mathcal{O}(1) $ in industrial systems. 
	In this way, ETA is able to reduce the complexity of similarity comparison between target item $ i $ and all behavior items $ i \in \mathcal{X}^m_{lt} $ from $\mathcal{O}(L \cdot d_k)$ by standard TA to $\mathcal{O}(L)$.
	
	After computing the hamming distance between target item and all behavior items, ETA-Net will perform MHTA between the target item and the selected top-$ K $ behavior items to model user's long-term interest. 
	In this way, the complexity of modeling long sequence based on standard TA can be reduced from $\mathcal{O}(L \cdot d_k)$ to $\mathcal{O}(K \cdot d_k)$ with $K \ll L$.  
	
	\subsubsection{End-to-end Learning.}
	\label{sec:end-to-end-learning}
	Recall that in the two-stage methods, the retrieval model and the CTR model are trained separately, such that the user and item representations used in different stages are not updated consistently. This mismatch may lead to inaccurate retrieval of the top-$k$ relevant items thereby harming the final performance.
	In contrast, our proposed ETA-Net can be trained in an end-to-end manner. 
	Specifically, the hashing function for the retrieval of similar items in ETA does not involve any trainable parameters, while the parameters within the MHTA structure can be trained with the base model jointly.
	Note that during the training of ETA, the fingerprints of the target item $ \bm{Q} $ and the behavior items $ \bm{K} $ are updated consistently with their embedding vectors.
	Therefore, ETA-Net can always use the newest fingerprints to retrieve the top-$ k $ most relevant items from the long behavior sequence so as to eliminate the aforementioned mismatch within the two-stage models for performance enhancement.
	The experimental results presented in Sec.~\ref{sec:experiment} also verify that such end-to-end learning capability can improve CTR prediction performance considerably in terms of both offline and online experiments.
	
	\begin{figure}[!tb]
		\centering
		\includegraphics[trim = 18 18 18 15, clip, width=7.5cm]{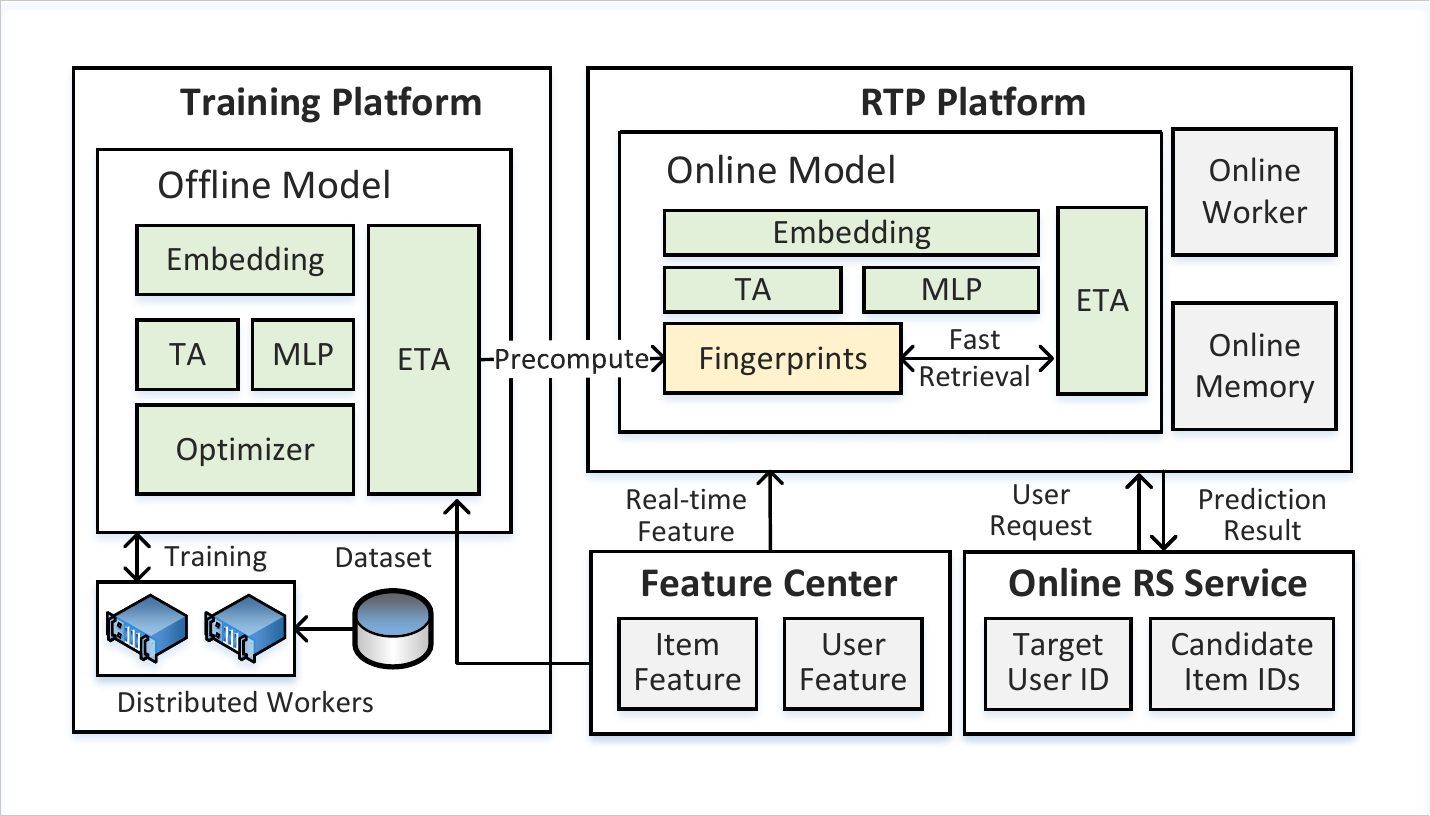}
		\caption{System Architecture for Online Deployment}
		\label{fig:architecture}
	\end{figure}
	
	\subsection{Online Implementation}
	\label{sec:architecture}
	In this section, we introduce the online implementation of ETA-Net in the RS of Taobao.
	The system architecture is able to handle one million user requests per second at traffic peak and respond within $ 20 $ milliseconds in general. It now serves the main traffic of Taobao to provide services to hundreds of millions of users towards billions of items in Taobao every day and brings 6.33\% lift on CTR and 9.7\% lift on GMV in average.
	
	\subsubsection{System Architecture}
	\label{sec:system_architecture}
	The architecture to implement our proposed ETA-Net in Taobao is presented in Fig.~\ref{fig:architecture} which describes the workflow of both offline training and online serving. 
	Generally, the offline training is based on a distributed machine learning platform. The trained ETA-Net will be uploaded to the Real Time Prediction~(RTP) platform for online serving. The RTP platform is able to retrieve user and item features from the feature center in real-time and feed them into ETA-Net for online inference.  
	Noticeably, the following two strategies can be adopted to reduce both computational and memory complexity of ETA-Net during the online inference process.
	
	The first one is to pre-compute the fingerprints of all items.
	Specifically, the model parameter are usually fixed during online inference in industrial systems in order to ensure system stability for online services.
	As such, it is viable to pre-compute the fingerprints of all items by performing SimHash~(which is parameter-free) over the offline well-trained model and keep the results in a embedding lookup table in advance.
	Then, during the online inference process, ETA-Net can directly make use of the pre-computed fingerprints from the embedding lookup table instead of computing the fingerprints in real-time so as to reduce the online inference complexity.
	Note that this precomputing strategy is different with the one used in SIM: the precomputed fingerprints in ETA-Net are still obtained from an end-to-end-trained CTR model, while the precomputed embeddings in SIM are obtained from an independently-trained retrieval model.

	The second one is to use integers to represent the binary fingerprints.
	Specifically, we can use an int64 integer to represent a binary fingerprint with $ 64 $-bit length. This trick can 1) save massive amount of offline storage when storing the precomputed fingerprints and 2) save lots of online machine memory when using fingerprints to retrieve items.
	
	\subsubsection{Complexity Analysis} 
	\begin{table*}[]
		\centering
		\caption{Comparison of the online model inference complexity. Here $ N_c $ is the number of target items to rank per user request; $ L $ is the number of behavior items within the long behavior sequence; $ K $ is the number of top-$ K $ behavior items retrieved from the long behavior sequence; $ d $ is the dimension of the embedding vector and $ d_p $ is the dimension of the projected embedding vector; $ M $ is the number of keys (i.e., item categories) used for subsequence retrieval within the second stage of SIM-hard. In real-world applications, these parameters usually scale as $ N_c=10^3, L=10^3, d_p=10^2, d=10^2, M=10^2, K=10^1 $. }
		\scalebox{0.85}{
			\begin{tabular}{@{}llll@{}}
				\toprule
				Long Sequence Model & Complexity of $ \bm{Q}$, $ \bm{K}$, $ \bm{V}$ & Complexity of $ \bm{A} = \bm{Q}\bm{K}^T $ & Complexity of $ \bm{A}\bm{V}^T $  \\ \midrule
				Standard TA    & $\mathcal{O}(N_c \cdot d \cdot d_p + L \cdot d \cdot d_p)$    & $\mathcal{O}(N_c \cdot L \cdot d_p)$   & $\mathcal{O}(N_c \cdot L \cdot d_p) $   \\
				Sim-hard (Stage-2)       & $\mathcal{O}(N_c \cdot d \cdot d_p + M \cdot K \cdot d \cdot d_p)$    & $\mathcal{O}(N_c \cdot K \cdot d_p)$   & $\mathcal{O}(N_c \cdot K \cdot d_p) $ \\
				Sim-soft (Stage-2)      & $\mathcal{O}(N_c \cdot d \cdot d_p + N_c \cdot K \cdot d \cdot d_p)$    & $\mathcal{O}(N_c \cdot K \cdot d_p)$   & $\mathcal{O}(N_c \cdot K \cdot d_p) $ \\
				ETA            & $\mathcal{O}(N_c \cdot d \cdot d_p + L \cdot d \cdot d_p + N_c \cdot d_p \cdot d_p + L \cdot d_p \cdot d_p)$    & $\mathcal{O}(N_c \cdot L + N_c \cdot K \cdot d_p)$   & $\mathcal{O}(N_c \cdot K \cdot d_p) $ \\
				ETA+           & $\mathcal{O}(N_c \cdot d \cdot d_p + L \cdot d \cdot d_p)$    & $\mathcal{O}(N_c \cdot L + N_c \cdot K \cdot d_p)$   & $\mathcal{O}(N_c \cdot K \cdot d_p) $ \\ \bottomrule
		\end{tabular}}
		\label{table:inference_complexity}
	\end{table*}
	
	\begin{table}[]
		\centering
		\caption{Comparison of the complexity of sequence retrieval from feature center.}
		\scalebox{0.85}{
			\begin{tabular}{@{}ll@{}}
				\toprule
				Long Sequence Model & Retrieval Complexity \\ \midrule
				Standard TA & $\mathcal{O}(1)$ \\
				Sim-Hard~(Stage-1) & $\mathcal{O}(M)$ \\
				Sim-Soft~(Stage-1) & $\mathcal{O}(N_c \cdot log(L))$ \\
				ETA & $\mathcal{O}(1)$ \\
				ETA+ & $\mathcal{O}(1)$ \\ \bottomrule
		\end{tabular}}
		\label{table:retrieval_complexity}
	\end{table}
		
	We compare the online inference complexity with the standard TA and SIM which is the SOTA solution for long sequence modeling in industrial systems in Table~\ref{table:inference_complexity}.
	Specifically, SIM proposed two methods to retrieve the top-$ K $ relevant items: 1) the soft-search method, which uses a trainable retrieval model to generate user/item embeddings to build retrieval indexes; and 2) the hard-search method, which directly uses user ID and item category to build retrieval indexes. We refer to them as Sim-hard and Sim-soft respectively in sequel.
	The inference complexity of TA mainly comes from three components: 
	1)~the computation of $ \bm{Q} \!=\! \bm{E}^t(\bm{W}_i^Q)^T $, $ \bm{K} \!=\! \bm{E}^s(\bm{W}_i^K)^T $ and $ \bm{V} \!=\! \bm{E}^s(\bm{W}_i^V)^T $;
	2)~the computation of $ \bm{A} = \bm{Q}\bm{K}^T $;
	and 3)~the computation of $ \bm{A}\bm{V}^T $.
	We compare the complexity of the above three components in sequel.

	For the first component, the recommendation model needs to evaluate the click probability of a list of $ N_c $ target items in order to select the top items per user request in online RS.
	Hence, $ Q $ needs to be calculated $ N_c $ times to obtain the projected feature of the $ N_c $ target items, respectively. This  leads to a complexity of $ \mathcal{O}(N_c \cdot d \cdot d_p) $ with $ d $ and $ d_p $ denoting the dimension of the original and projected embedding vector, respectively. 
	On the other hand, the user features~(e.g., the long behavior sequences) are usually fixed\footnote{The user features are updated~(every few seconds or every few hours) much slower than the occurrence of user requests~(every few milliseconds).} during each user request in online RS. Therefore, both $ K $ and $ V $ only need to be updated once per user request. 
	
	Both SIM-hard and SIM-soft need to perform feature crossing between $ \bm{Q} $ and $ \bm{K}, \bm{V} $ in order to select subsequences from the long behavior sequence at the first stage. This leads to a complexity of $\mathcal{O}( M \cdot K \cdot d \cdot d_p)$ for SIM-hard and $\mathcal{O}( N_c \cdot K \cdot d \cdot d_p)$ for SIM-soft, where $ M $ refers to the number of keys (i.e., item categories) used for subsequence retrieval. 
	In contrast, the item retrieval in ETA-Net relies on the comparison of fingerprints of the behavior items. 
	The generation of the fingerprints of $ \bm{Q} $ and $ \bm{K} $ relies on the SimHash algorithm, which leads to a complexity of $ \mathcal{O}(N_c \cdot d_p \cdot d_p) $ and $ \mathcal{O}(L \cdot d_p \cdot d_p) $, respectively. 
	Moreover, the fingerprints can be pre-generated as stated in Sec.~\ref{sec:system_architecture}. This can further reduce the complexity of ETA-Net to $\mathcal{O}(N_c \cdot d \cdot d_p + L \cdot d \cdot d_p)$, which makes the complexity of ETA-Net similar as standard TA and SIM-hard but much smaller than SIM-soft. 
	We refer to the ETA with and without the fingerprint pre-generation strategy as ETA+ and ETA in Table~\ref{table:inference_complexity}.
	
	For the second and third component, both ETA-Net and SIM only needs to perform feature crossing among the top-$ K $ behavior items extracted from the long behavior sequence. This reduces the complexity from $\mathcal{O}(N_c \cdot L \cdot d_p) $ of standard TA to $ \mathcal{O}(N_c \cdot K \cdot d_p) $ where $ K \ll L $. Meanwhile, as aforementioned in Sec.~\ref{sec:ETA}, ETA needs to measure the similarity of $ \bm{Q} $ and $ \bm{K} $ for item retrieval by comparing their hamming distance through bit-wise $ xor $ operation. The $ xor $ operation incurs constant complexity in industrial systems such that the complexity of similarity measurement in ETA-Net is $\mathcal{O}(N_c \cdot L)$, which is small compared with $\mathcal{O}(N_c \cdot K \cdot d_p) $.
	
	Additionally, we also compare the complexity of sequence retrieval from the feature center in Table~\ref{table:retrieval_complexity}, which is also costly in online RS. Specifically, in online RS, the long behavior sequence needs to be transmitted from the feature center to the deployment platform~(i.e., RTP platform as shown in Fig.~\ref{fig:architecture}) for model inference per user request. 
	Both standard TA and ETA take the complete long behavior sequence as the input, such that the feature center only needs to retrieve the complete long behavior sequence \textit{once} based on an inverted index structure~(long behavior sequences are indexed according to the user IDs in a one-to-one manner) in $ \mathcal{O}(1) $.
	In SIM-hard, the long behavior sequence needs to be split into multiple subsequences based on item categories. As such, per user request, the feature center needs to retrieve $ M $ subsequences for one target user, which leads to a complexity of $ \mathcal{O}(M) $ where $ M $ denotes the number of categories appeared in the list of target items.
	In SIM-soft, the retrieval complexity is higher due to the use of a trainable retrieval model. The retrieval complexity of all target items can be reduced to $ \mathcal{O}(N_c \cdot log(L)) $ by using ALSH method~\cite{qi2020search}. 
	
	\section{Experiments}
	\label{sec:experiment}
	In this section, we conduct extensive experiments on both offline datasets and online RS with the goal to answer the following research questions.
	
	\smallskip\noindent\textbf{Q1:} Does ETA outperform the SOTA recommendation methods?
	
	\smallskip\noindent\textbf{Q2:} How efficient is ETA compared with other methods?
	
	\smallskip\noindent\textbf{Q3:} Which part of ETA contributes the most to the prediction performance or efficiency? 
	
	\subsection{Experimental Setup}
	
	\subsubsection{Datasets.}
	We use one public benchmark dataset and one industrial dataset collected from the online RS in Taobao for offline evaluation. The statistics is given in Table~\ref{tab: ranking}.
	
	\textbf{Taobao dataset}\footnote{\url{https://tianchi.aliyun.com/dataset/dataDetail?dataId=649&userId=1}} released by~\cite{zhu2019joint} is a widely used public benchmark dataset for CTR predictions~\cite{qi2020search, qin2020user}, which contains over $ 100 $ million instances collected from Taobao Mobile App. The user behaviors include click, favorite, add to cart and purchase. Each user has around $ 101 $ recent behaviors in average. We choose the recent $ 16 $ behaviors as the short-term user behaviors and the recent $ 256 $ behaviors as the long-term user behaviors. Since the Taobao dataset only contains positive interactions, i.e., click, favorite, add to cart and purchase, we follow MIMN~\cite{pi2019practice} to select users' last interacted item as the positive sample and randomly sample a batch of items within the same category as the positive sample to be the negative samples.
	The dataset is split into training set (80\%), validation set (10\%) and test set (10\%) according to the timestamps.

	\textbf{Industrial dataset}~\footnote{This dataset will be released to public to assist future research on long sequential data modeling.} is collected from one of the largest recommendation application in Taobao, involving billion scale of users and items. The industrial dataset contains both positive and negative interactions~(e.g., impression without user clicks) such that negative sampling is not needed. There are over $ 142 $ billion instances and each user has around $ 938 $ recent behaviors in average, which is much longer than the sequences from the public dataset. We choose the recent $ 48 $ behaviors as the short-term user behaviors and the recent $ 1024 $ behaviors as the long-term user behaviors. We also compare the use of $256$, $512$ and $2048$ as long-term user behaviors in the ablation study. Following SIM~\cite{qi2020search}, we use the instances of past two weeks' as the training set and the instances of the next day as the test set.
	
	\begin{table}[t]
		\centering
		\caption{Statistics of the datasets used in this paper.}
		\label{tab:freq}
		\scalebox{0.8}{
			\begin{tabular}{ccccc}
				\toprule
				& Users & Items &  Categories & Instances \\
				\midrule
				Taobao & 987,994 & 4,162,024 &  9,439 & 100,150,807 \\
				Industrial & 0.4 billion & 0.7 billion & 24,568 & 142 billion \\
				\bottomrule
		\end{tabular}}
		\label{tab: ranking}
	\end{table}
	
	\subsubsection{Compared Methods.}
	\label{sec:comaring_method}
	We compare our model with the following mainstream baselines for CTR prediction.
	
	\begin{itemize}
		\item \textbf{Pooling-based DNN}: This model replaces the TA structure in DIN with average pooling to make use of user behavior sequences, similar as the YouTube model~\cite{covington2016deep}. 
		\item \textbf{DIN}~\cite{zhou2018deep}: This is a widely used benchmark for sequential user data modeling in CTR predictions, which models short  behavior sequences with TA. 
		\item \textbf{DIN-Long}: This model uses pooling-based DNN to model long behavior sequences while using TA to model short behavior sequences, which is the simplest way to add long-term interests to DIN.
		\item \textbf{UBR4CTR}~\cite{qin2020user}: This is a recently proposed two-stage long sequence model. In UBR4CTR, the query is generated by a feature selection model for item retrieval. UBR4CTR has not been evaluated in online RSs yet.
		\item \textbf{SIM}~\cite{qi2020search}: This is the SOTA industrial solution for long sequence modeling based on the two-stage paradigm. SIM has been deployed in the display advertising system in Alibaba, serving the main traffic.
	\end{itemize}
	
	It is noteworthy that SIM surpasses MIMN in both offline experiments and online A/B tests under the same production environment~\cite{pi2019practice,qi2020search}, such that we only compare our proposed ETA-Net against SIM in this paper. 
	As aforementioned, SIM proposed two methods to retrieve the top-$ K $ relevant items: 1) the soft-search method, which uses a trainable retrieval model to generate user/item embeddings to build retrieval indexes; and 2) the hard-search method, which directly uses user ID and item category to build retrieval indexes. SIM chooses the hard-search method as the final solution for online deployment since the soft-search method requires massive resource consumption. 
	In this paper, we choose SIM with hard-search as the comparing method since it is the SOTA industrial solution and also serves the main traffic of Taobao. We denote SIM-hard as SIM for short in sequel.
	
	\subsubsection{Metrics.}
	We compare the performance of different methods using the metrics of Area Under ROC~(AUC), CLICK, GMV and CTR.
	Specifically, AUC is a widely used metric for binary classification in CTR prediction, we use AUC for offline evaluation and use CLICK and CTR for online A/B tests. Here, CLICK refers to the total number of clicked items, while CTR is defined as \mbox{CLICK/PV} with PV denoting the total number of impressed items. CTR measures users' willingness to click and is therefore a widely used metric in industrial applications. GMV is a term used in online retailing to indicate a total sales monetary-value for merchandise sold over a certain period of time.
	We compare the efficiency of different methods with their inference time. Inference time is defined as the total time cost of predicting the CTR scores of a batch of items per user request. It is measured by deploying the trained models on industrial platforms, i.e., RTP in this paper, and recording the averaged time cost per user request. In our experiments, we evaluate the inference time of different methods under the same production environment~(using the same number of machines and parameter settings, etc.) in order to give a fair comparison.
	
	\subsubsection{Parameter Settings.}
	\label{sec:parameter_setting}
	In all experiments, we use the validation set to tune the hyper-parameters to generate the best performance for different methods. The learning rate is searched from $10^{-4}$ to $10^{-2}$. The L2 regularization term is searched from $10^{-4}$ to $1$. All models use Adam~\cite{kingma2014adam} as the optimizer. The batch size is set to be $ 256 $ and $ 1024 $ for Taobao and the industrial dataset, respectively. 
	
	\begin{table}[]
		\centering
		\caption{Experimental results on offline evaluations.}
		\scalebox{0.85}{
			\begin{tabular}{@{}lccc@{}}
				\toprule
				& Taobao Dataset         & \multicolumn{2}{c}{Industrial Dataset} \\ \cmidrule(l){2-4} 
				Method                    & AUC             & AUC              & Inf. Time   \\ \midrule
				Pooling-based DNN         & 0.8442          & 0.7216           & 8ms          \\
				DIN                       & 0.8626          & 0.7279           & 11ms          \\
				DIN-Long                  & 0.8661          & 0.7311           & 14ms          \\
				UBR4CTR                   & 0.8651          & 0.7318           & 41ms          \\
				UBR4CTR+timeinfo          & 0.8683          & 0.7331           & 41ms          \\
				SIM                       & 0.8675          & 0.7327           & 21ms          \\
				SIM+timeinfo              & 0.8708          & 0.7338           & 21ms          \\
				ETA-Net                   & 0.8721          & 0.7361           & 19ms          \\
				\textbf{ETA-Net+timeinfo} & \textbf{0.8746} & \textbf{0.7373}  & \textbf{19ms} \\ \bottomrule
		\end{tabular}}
		\label{tab:offline}
	\end{table}
	
	\setlength{\tabcolsep}{10pt}
	\begin{table}[t]
		\centering
		\caption{Experimental results on online A/B tests. }
		\scalebox{0.8}{
			\begin{tabular}{lccc}
				\toprule
				Method  & CTR & GMV & Inf. Time \\
				\midrule
				SIM+timeinfo & 4.53$\%$ & 6.6$\%$ & 21ms \\
				ETA-Net+timeinfo& 6.33$\%$ & 9.7$\%$ & 19ms   \\
				\bottomrule
		\end{tabular}}
		\label{tab:online}
	\end{table}

	\subsection{Offline Evaluation}
	\subsubsection{Results on Public Dataset.} 
	The AUC improvements on Taobao dataset are presents in Table~\ref{tab:offline}. The results verify that our proposed ETA-Net can outperform all the other comparing methods. 
	Specifically, ETA-Net outperforms SIM by $ 0.0046 $ and outperforms DIN with long behavior sequences by $ 0.006 $. Besides, we follow SIM~\cite{qi2020search} to add the time embeddings to each method and compare their performances. The results show that our proposed ETA-Net still outperforms SIM by $ 0.0038 $ and outperforms DIN with long behavior sequences by $ 0.0085 $. 
	The results also show that 1) DIN with long behavior sequences outperforms DIN by $ 0.0035 $; and 2) DIN outperforms Pooling-based DNN by $ 0.0184 $, which indicates that 1) it is beneficial to model long behavior sequences in CTR model; and 2) TA is much superior than average pooling in modeling sequential data.
	We also find that UBR4CTR performs worse than DIN with long sequence. Recall that UBR4CTR uses a feature selection model to select the behavior items whose features(eg. category, weekday) are the same with the target item. However, although this filtering can help remove certain noise, it will also leads to much shorter sequence length: the average sequence length after the filtering in UBR4CTR is $ 4.8 $, while the average sequence length in DIN is $ 86.4 $.
	
	\subsubsection{Results on Industrial Dataset.}
	The AUC improvements on industrial dataset are presents in Table~\ref{tab:offline}. 
	It is worth mentioning that $ 0.001 $ improvement of AUC can bring millions of real-world clicks thereby leading to considerable improvements of revenue in our online RS. In this context, our proposed ETA-Net outperforms SIM by $ 0.0034 $ and outperforms UBR4CTR by $ 0.0043 $.
	Moreover, after adding the time embeddings, ETA-Net outperforms SIM by $ 0.0035 $ and outperforms UBR4CTR by $ 0.0042 $. These improvements are remarkable in our RS.
	
	We also measure the inference time of different methods, which decides their applicability to industrial systems. The results are shown in Table~\ref{tab:offline}. 
	It is remarkable that the inference time of our proposed ETA-Net is even less than which of SIM. Considering that ETA-Net enables end-to-end learning of long sequence retrieval while SIM retrieves items using a non-parametric method based on user ID and item category, this improvement on inference efficiency verify that the SimHash-based ETA and the dedicated architecture presented in Sec.~\ref{sec:architecture} are highly efficient in industrial systems.  
	On the other hand, UBR4CTR has the maximum inference time due to the use of an extra feature selection model before the retrieval stage. It also requires an IDF and BM25 based procedure to retrieve top-$k$ items in online system, which is time-consuming. The inference time of UBR4CTR may hinder its application to large-scale recommendation tasks.
	
	\subsubsection{Ablation Study.} 
	\begin{figure}[tb]
		\centering
		\includegraphics[trim = 1 1 1 1, clip, width=7cm]{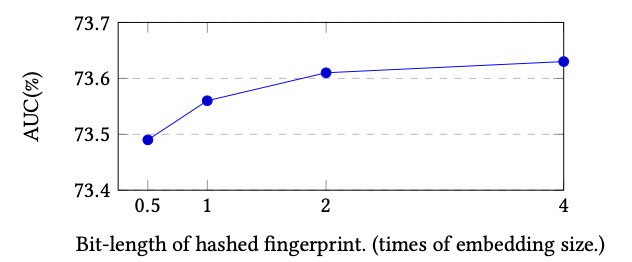}
		\caption{Ablation study of the bit-length in ETA-Net on the industrial dataset. }
		\vskip -1em
		\label{fig:fingerprints}
	\end{figure}
	We investigate the performance of ETA-Net using different encoding strategy, aiming at yielding insights behind long sequence modeling to answer RQ3. The results on the industrial dataset are shown in Table~~\ref{table:ablation}. Specifically, we compare the encoding strategy from the following aspects: 1) the modeling technique for sequential data, i.e., TA or average pooling~(AVG); 2) the length of the long sequence; 3) the number of items to retrieve, i.e., the $ K $ in top-$ K $ retrieval; 4) the item retrieval technique, i.e., SimHash~(HASH) or dot-product~(DOT). For example, TA/HASH/48/1024 in Table~\ref{table:ablation} means we use TA to model sequential data and use SimHash to retrieve top-$ 48 $ items from a long sequence with length $ 1024 $. 
	
	We observe that
	1) the version V4 uses TA to directly model the entire $ 1024 $ behavior items can achieve the best performance, at the cost of the highest inference time. On the other hand, the version V0 only uses $ 48 $ behavior items and can reduce the inference time by $ 46\% $ with only $ 0.001 $ performance loss on AUC;
	2) the difference between the version V3 and the version v0 shows that replacing the SimHash with dot product to retrieve items can achieve $ 0.0007 $ improvement on AUC. However, the inference time also increases by $ 68\% $, which is not cost-effective in online system; 
	and 3) there are trade-offs between AUC and inference time as indicated by the difference among version V2.X versus version V0, such that it is necessary to select an appropriate sequence length according to actual demands in order to obtain the optimal cost-efficiency in online systems.
	
	Moreover, we also evaluate the performance of ETA-Net when using different number of bits to generate the fingerprints. The results are shown in Figure~\ref{fig:fingerprints}. 
	We observe that 1) AUC can be improved by increasing the bit-length of $\bm{h}$; 2) when the bit length is larger than twice of the embedding size, the improvement on AUC becomes marginal. 
	
	\begin{table}[]
		\centering
		\caption{Ablation study of the encoding strategies in ETA-Net on the industrial dataset.}
		\scalebox{0.8}{
			\begin{tabular}{@{}llcc@{}}
				\toprule
				Version & Encoding Strategy & AUC    & Inf. Time \\ \midrule
				V0           & TA/HASH/1024/48   & 0.7361 & 19ms                 \\ 
				V1           & AVG/-/1024/-      & 0.7311 & 14ms                 \\ 
				V2.1         & TA/HASH/256/48    & 0.7339 & 14ms                 \\
				V2.2         & TA/HASH/512/48    & 0.7348 & 16ms                 \\
				V2.3         & TA/HASH/2048/48   & 0.7394 & 23ms                 \\ 
				V3           & TA/DOT/1024/48    & 0.7368 & 32ms                 \\ 
				V4           & TA/-/1024/-       & 0.7371 & 35ms                 \\ \bottomrule
		\end{tabular}}
		\vskip -1em
		\label{table:ablation}
	\end{table}
	
	\subsection{Online Evaluation}
	\label{sec:ablation}
	Since 2021, ETA-Net has been deployed in Taobao based on the proposed system architecture to serve a bunch of main recommendation scenarios, e.g., Taobao Guess-you-like at the front page and the post-buy recommendation page presented after user payment. The platform serves more than $ 120,000 $ QPS at a traffic peak and provides recommendation service to hundreds of millions of users toward billions of items every day.
	
	The online experimental results of online A/B test are shown in Table~\ref{tab:online}, which are collected from 2021-01-05 to 2021-02-05. 
	The improvements are evaluated by comparing the methods with a DIN-like model which uses standard TA to model short behavior sequences without long behavior sequences~(our previous main traffic model). 
	The results show that our proposed ETA-Net obtains $ 6.33\% $ improvement on CTR and $ 9.7\% $ improvement on GMV, which verifies the effectiveness and necessity to model long behavior sequences in industrial applications.
	Moreover, we also implement the SOTA industrial solution SIM in the platform and compare its performance with ETA-Net. The A/B test verifies that our proposed ETA-Net also outperforms SIM by $ 1.8\% $ in terms of CTR and $ 3.1\% $ in terms of GMV. 
	Note that $ 1\% $ improvement on CTR or GMV is a considerable improvement in real-world RS, especially for applications with billion-scale user and items. For example, in scenarios such as Taobao Guess-you-like, $ 1\% $ improvement on CTR and GMV can bring millions of clicks and tens of millions amount of transactions every day.
	Additionally, our solution also outperform SIM in terms of the inference time (i.e., $ 19 $ms v.s. $ 21 $ms), which is remarkable since ETA-Net is an end-to-end model.
	The online experimental results verify that our proposed ETA-Net is able to model long-term user interest in a more effective and efficient manner.
	
	\begin{table}[]
		\centering
		\caption{Gain of long-term user interest modeling. }
		\scalebox{0.8}{
			\begin{tabular}{@{}cccc@{}}
				\toprule
				& Short-term (0-5d) & Mid-term (6-50d) & Long-term (50d) \\ \midrule
				Click & 3.12\%            & 6.58\%           & 13.49\%         \\
				GMV   & 5.18\%            & 9.2\%            & 16.68           \\ \bottomrule
		\end{tabular}}
		\vskip -1em
		\label{table:gain}
	\end{table}
	We also present the performance gain of introducing long-term user behavior sequence into the online product model in Table~\ref{table:gain}. 
	Specifically, for each user, we split his/her interacted item categories into three types: 1)~have NOT clicked/purchased within a short-term time period~(i.e., less than recent $ 5 $ days), 2)~have NOT clicked/purchased within a mid-term time period~(i.e., more than recent $ 6 $ days but less than recent $ 50 $ days) and 3)~have NOT clicked/purchased within a long-term time period~(i.e., more than $ 50 $ days). 
	We compare the performance of ETA-Net with which of our DIN-like baseline over the above three types of categories and average the results among all the users to give an overview. 
	The result in Table~\ref{table:gain} shows that ETA-Net generates superior performance over all three types of categories. Noticeably, the gain is more considerable over the third type. This verifies that ETA-Net is able to awake users' long-term interest by recommending items according to their long-term behaviors.
	
	\section{Conclusion}
	In this paper, we propose the ETA-Net for CTR prediction. To the best of our knowledge, ETA-Net is the first method that is able to model the long sequential user data in an end-to-end manner and is of high cost-efficiency on large-scale recommendation tasks.
	Moreover, we also present a general architecture to implement the proposed ETA-Net on industrial systems. 
	The experiments on both offline and online evaluation verify the superiority of ETA-Net against other SOTA models. 
	It is noteworthy that ETA-Net has been implemented in the RS of Taobao to serve the main traffic and bring 6.33\% lift on CTR and 9.7\% lift on GMV to bring millions of clicks and tens of millions amount of transactions every day.
	
	\bibliographystyle{ACM-Reference-Format}
	\bibliography{DLPKDD_ETA}
	
\end{document}